# SEMANTIC SNIPPET CONSTRUCTION FOR SEARCH ENGINE RESULTS BASED ON SEGMENT EVALUATION

K.S.Kuppusamy[1] & G.Aghila[2]

The result listing from search engines includes a link and a snippet from the web page for each result item. The snippet in the result listing plays a vital role in assisting the user to click on it. This paper proposes a novel approach to construct the snippets based on a semantic evaluation of the segments in the page. The target segment(s) is/are identified by applying a model to evaluate segments present in the page and selecting the segments with top scores. The proposed model makes the user judgment to click on a result item easier since the snippet is constructed semantically after a critical evaluation based on multiple factors. A prototype implementation of the proposed model confirms the empirical validation.

## 1. Introduction

The mammoth amount of pages in the World Wide Web has made the search engines, an inevitable tool for the web users. Web pages are getting visited through the search engine result lists rather than being directly entering their URL in to the browser. The judgment of the user whether to click a link or not, from the search result listing is one of the important parameters that define the user count for a web page.

The result listing of a search engine generally consists of a link to the page and a snippet associated with it. The snippets hold the occurrences of the user specified keywords in the target page. The length of the terms to be gathered is determined by the window size. In the popular Google search engine, a snippet size of 100 characters is been used [1]. This paper addresses the snippet construction problem so that the constructed segment would reflect the page's relevancy with respect to the user supplied query. The objectives of this research work are as listed below:

- Proposing a model to construct the search result snippets by evaluating the relevancy of the query with respect to the page based on various semantics.

- Analyzing the effectiveness of the snippet construction in making the user to reach the required information quickly and efficiently.

The remainder of this paper is organized as follows: Section 2 list out various research works carried out related to the proposed model. Section 3 would elaborate the proposed model and the algorithms. Section 4 illustrates the prototype implementation and analysis of experimental results. Section 5 is about conclusions and future directions for this research work.

## 2. Related Works

The web page segmentation has been explored by various researchers. There exist various approaches to segment a web page. Cao et al [2] has proposed a segmentation method based on image processing techniques. The web page is considered as an image and features are used to segment the web page. Kohlschütter et al [3] has applied the text density of sections of the page to carry out the segmentation process. A graph theory based segmentation approach is explained by Deepayan Chakrabarti et al [4].

The Vision based Page segmentation (VIPS) process explained by D.Cai et al [5] provides the segmentation approach based on visual features. Since this approach is closer to the human perusal of a web page, the segmentation for this work has been carried out with this approach.

Learning the importance of the page segments is explored by Ruihua Song et al [6]. It has been indicated that differentiating noisy segments from the informative ones can be a very handy mechanism in better managing the web page for mining and other processes. Lin et al [7] has explored a table based approach to identify informative portions from a web page. A unified approach to document similarity search using manifold-ranking of blocks has been proposed by Xiaojun Wan et al [8]. In [9], we have proposed a model called Museum to evaluate a segment based on multiple factors. Segmentation based dynamic page construction from search engine results are explained in [10]. A semantically driven segment selection for focussed search is explained in [11].

[1, 2]Department of Computer Science, School of Engineering and Technology, Pondicherry University, Pondicherry, INDIA
E-mail: [1]kskuppu@gmail.com, [2]aghilaa@yahoo.com

## 3. The Model

The proposed model facilitates the construction of search result snippet by semantically evaluating the web page in a fine-grained manner.

$$\Upsilon = \left[\frac{I}{Q}\right]^{\mu} \quad (1)$$

In (1), the index is denoted as I and the user supplied query is represented as Q. The $\mu$ represent the threshold limit up to which the resultant items would be considered for processing. Upon execution of the query, (1) can be expanded as shown in (2).

$$\Upsilon = \left[\frac{I}{Q}\right]^{\mu} = \{\Gamma_1, \Gamma_2, ..., \Gamma_{\mu}\} \quad (2)$$

In (2), each $\Gamma_i$ represents the resultant pages. Each of these pages can be segmented as shown in (3).

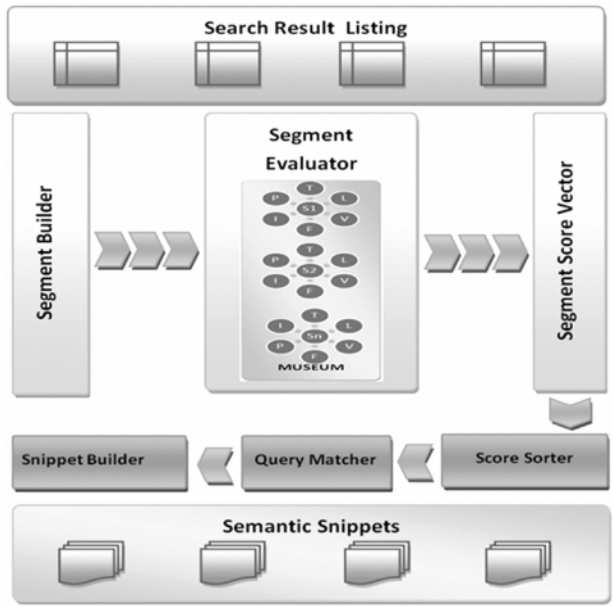

Fig.1: Semantic Snippet Construction

$$\Gamma_i = \bigcup_{j=1}^{n} \alpha_j \quad (3)$$

The evaluation of these segments would be done by the Museum model. It performs the multi-dimensional evaluation of each of the segments as shown in (4).

$$\omega(\alpha_j) = (F, E, L, V, R, M) \quad (4)$$

Each segment would be semantically evaluated as sum of six different weights co-efficient. In(4) F denotes Freshness, E is Theme, L stands for Link, V indicates Visual, R denotes Profile, and M is Image. The final score would be the sum of all these scores. The detailed procedures to calculate these weights are as shown in [9].

$$\Gamma_i = \bigcup_{j=1}^{n} \alpha_j = \sum_{k=1}^{m} |F_k + E_k + L_k + V_k + R_k + M_k| \quad (5)$$

The result of the summation process in (5) would yield a vector consisting of weighted values for each segment in a page. This vector would be sorted in the descending order of weights as shown in (6).

$$\Gamma_i = \{\langle\omega(\alpha_1)\rangle, \langle\omega(\alpha_2)\rangle ... \langle\omega(\alpha_k)\rangle \\ \forall_{m=1}^{k}[\omega(\alpha_{m-1}) \geq \omega(\alpha_m)]\} \quad (6)$$

For the snippet construction process the occurrence of keywords can be selected from the top scored segments as shown in (7).

$$Snip(\Gamma_i) = \bigcup_{j=1}^{m}[match(\alpha_j, Q)] \quad (7)$$

The algorithm to implement the proposed model is as shown below:

```
Algorithm SemanticSnippet
Input   : Result List Υ, Query Q
Output  : Constructed snippets
Begin
   for each result page Γ_i in Υ
   begin
      //segment the page
      Γ_i = buildsegment();
      for each segment α_j in Γ_i
      begin
         //calculate segment using Museum model
         Wgt (α_i) = Museum (α_j)
      end
      //sort the segments in decreasing order of wgts
      Sort (wgt(α_i))
      fetch the query matching words from these segments.
      Snip [i] = match(Q, α_i)
   end
End
```

## 3. The Experimental Results

A prototype implementation of the proposed system is done on the Linux, Apache, Mysql and PHP stack. The segment evaluation is implemented as a library with the Museum model. The comparisons are made between the system with simple snippets and system with semantic snippets.

The experiments are conducted on the prototype implementation checking for the number of result items which can be explicitly judged with simple text snippets and semantic snippets. The results of a user session with fifteen searches are tabulated in Table 1.

Table 1
Number of Explicitly Judge-able Result Items

| Session | 1 | 2 | 3 | 4 | 5 | 6 | 7 | 8 | 9 | 10 | 11 | 12 | 13 | 14 | 15 |
|---|---|---|---|---|---|---|---|---|---|---|---|---|---|---|---|
| Sim. Snip | 2 | 3 | 4 | 5 | 1 | 7 | 3 | 5 | 6 | 0 | 6 | 1 | 5 | 2 | 1 |
| Sem. Snip | 8 | 7 | 9 | 8 | 10 | 8 | 9 | 9 | 10 | 7 | 9 | 6 | 8 | 7 | 8 |

The chart with comparison between simple snippets and semantic snippets are depicted in Fig.2.

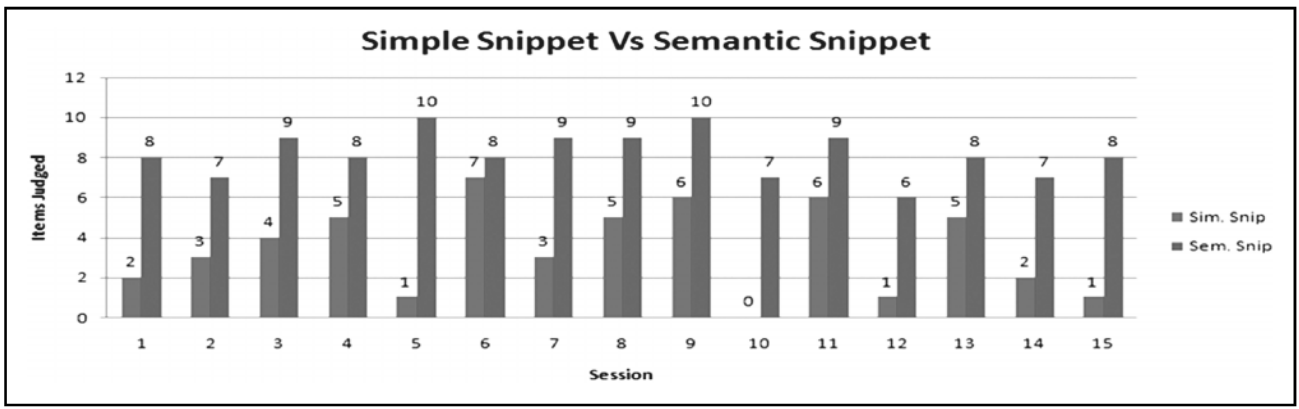

Fig.1: Simple Snippet Vs Semantic Snippet Relevancy Judging

## 5. Conclusions and Future Directions

The result of the experiments conducted on the prototype implementation concludes that the relevancy judgment of the result item can be improved with the incorporation of semantic snippets. In the simple snippet based approach the number of result item which can be judged for relevancy by the user is 3.4 and in the case of semantic snippets this metric raises to 8.2. The future directions for this research work include incorporation of images and other resources in the snippet which would further enhance the relevancy judgment process.